# Spatiotemporal Topological Phase Transition in non-Hermitian Photonic System

Zimeng Zeng, Zhuoyang Li, Jiayao Liu, Zelong He, and Zhaona Wang*

Key Laboratory of Multiscale Spin Physics (Ministry of Education), Applied Optics Beijing Area Major Laboratory, School of Physics and Astronomy, Beijing Normal University, Beijing 100875, China

*Contact author: zhnwang@bnu.edu.cn

## Abstract

While energy-band topology in spatial photonic crystals (PCs) and momentum-band topology in temporal crystals have each served as powerful probes of topological phases in their respective domains, their unification in a static platform remains unexplored. In this Letter, we bridge this gap by introducing a waveguide assisted non-Hermitian SSH model, in which controlled tuning of loss and coupling drives PT-symmetry breaking and enables a continuous transition between energy- and momentum-gap regimes. This allows us to construct a complete spatiotemporal topological phase diagram in a unified parameter space. By mapping this phase diagram onto a spatially graded PC, we experimentally observe multiple Bloch momentum-band gaps and a continuous spatiotemporal topological transition via translating across the static sample, enabling real-time control over the evolution pathway of the band topology. Our work creates a versatile, bias-free platform for exploring synthetic spacetime physics and opens new avenues for controlling light via non-Hermitian band engineering.

In condensed matter system and photonics, symmetry breaking induced by periodicity constitutes the emergence of a wealth of novel states[1,2]. Periodicity in spatial photonic crystals (PCs) gives rise to energy-bands, whose topological properties are commonly modulated by the closing and reopening of energy-band gaps. Temporal periodicity, as realized in temporal PCs, leads to the formation of momentum-bands, whose topology is manifested in the evolution of momentum-band gaps[3-10]. While both frameworks rest on a common theoretical foundation, their topological consequences have largely been studied in isolation[11-13]. Inspired by the transition from the

Newtonian conception of absolute space and time to relativistic view of spacetime, PCs are now advancing beyond single dimensional topology toward coupled spatiotemporal topology. A central challenge in spatiotemporal topology is the need for physical temporal modulation of material parameters which is extremely demanding at optical frequencies[14]. To circumvent these optical-frequency limitations, recent efforts have shifted to microwave implementations based on the dynamic transmission lines composed of microstrips periodically loaded with varactor diodes and inductors[15].

Interestingly, recent studies succeeded in revealing the intrinsic connection between non-Hermiticity in spatial PCs and temporal topology, enabling the achievement of synthetic temporal PCs[16-21]. In non-Hermitian system, spatially modulated imaginary refractive index acts as an effective complex periodic potential, whose PT-symmetry breaking gives rise to exceptional points (EPs) and opens a momentum-band gap[22]. This breaking of time-reversal symmetry allows a static photonic structure to realize momentum-band topology previously associated with temporal crystals. Moreover, by switching the topological phase of Bloch momentum-bands, temporal topological boundary states have been successfully observed in non-Hermitian electrical[23] and acoustic platforms[24]. The spatiotemporal topology is further achieved via pulse sequence modulation described by a time-dependent effective Hamiltonian.[25]. However, researches based on this concept remain yoked to time-dependent framework even though Bloch momentum-band does not require temporal modulation. This raises a fundamental question: is it possible to achieve spatiotemporal topology by a completely static Hamiltonian? More fundamentally, a global phase diagram offering a predictive framework of spatiotemporal topological phases is needed.

Here, we propose and realize a static platform for spatiotemporal topology based on a waveguide assisted non-Hermitian Su-Schrieffer-Heeger (SSH) model. In this system, the waveguide enables the control of both relative loss and coupling strength. Their interplay directly governs PT-symmetry breaking and defines a two-dimensional parameter space, in which the band structure switches continuously between energy-band and Bloch momentum-band regimes with a special boundary. A Berry-phase analysis along an open path within this synthetic space reveals four distinct topological phases, enabling continuous spatiotemporal transitions without any temporal modulation (Fig. 1a). We map the parameter space onto a gradient spatial PC, where structural parameters vary continuously. Based on this platform, we demonstrate a continuous spatiotemporal topological transition. Along the gradient direction, the system switches between topologically trivial and nontrivial energy-band phases while crossing distinct momentum-band gap regions.

*Spatiotemporal topology phase map*— We start with a waveguide coupled one-dimensional spatial PC incorporating optical loss (Fig. 1b), whose duty cycle (DC) $D= w/\Lambda$ ($w$ is the ridge width and $\Lambda$ is the grating period) and etching depth (ED) $E$ jointly determine the coupling strength and loss of two modes[26]. This coupling can be described by a 1D non-Hermitian SSH model (Fig. 1c). To reveal its underlying symmetry structure, a constant background term is extracted from this scenario and turns it into a PT-symmetry form without influencing the topology of system (Supplemental Material S1). This leads to the two-dimensional parameter space spanned by relative loss and coupling strength, within which PT symmetry can be continuously tuned from unbroken to broken, giving rise to different spectral regimes (Fig. S1). In the PT-unbroken phase, the system exhibits a real energy-band gap. When PT symmetry breaks, a pair of EPs opens a Bloch momentum-band gap, across which the imaginary part of the eigenvalues splits into positive and negative branches (Fig. S1c and S1d) [27-30]. This splitting directly corresponds to the exponential growth or decay of light waves traversing such a gap in a temporal PC[23,25]. The simultaneous presence of an energy-band gap and a momentum-band gap suggests that the two-dimensional parameter space effectively serves as a proxy for both space and time. This mapping transforms the parameter space itself into a synthetic spacetime arena, enabling the study of spatiotemporal topology in a completely static setting.

However, the emergence of EPs, as well as the non-chiral and non-Hermitian nature of this Hamiltonian, renders conventional methods for analyzing topological invariants[31-37]. The inability to compute topological invariants does not signal a trivial system[38-43], but rather calls for new theoretical approaches. Here, a separate analysis of the momentum- and energy-bands is employed. Considering the dispersion relation in a small-momentum region (see details in Supplemental Material S2), an effective Hamiltonian can be written as:

$$H_{\delta\omega}(\delta k) = \begin{pmatrix} -\delta k - i\frac{\eta}{2} & m + i\lambda\delta k \\ m - i\lambda\delta k & \delta k + i\frac{\eta}{2} \end{pmatrix}, \quad (1)$$

where $\eta$ and $\lambda$ are the dimensionless relative loss and relative coupling strength, respectively, and $m$

= 1-λ. The two-dimensional parameter space (λ, η) is separated into four distinct regimes, by the phase boundary condition

$$\eta^2\lambda^2 - 4m^2(1+\lambda^2) = 0, \tag{2}$$

which is marked by the yellow dashed line in Fig. 2a. This spatial and temporal topological phase transition boundary is marked by a Dirac cone in the band structure, between which are the energy-band (blue regions) and momentum-band (purple regions) gaps.

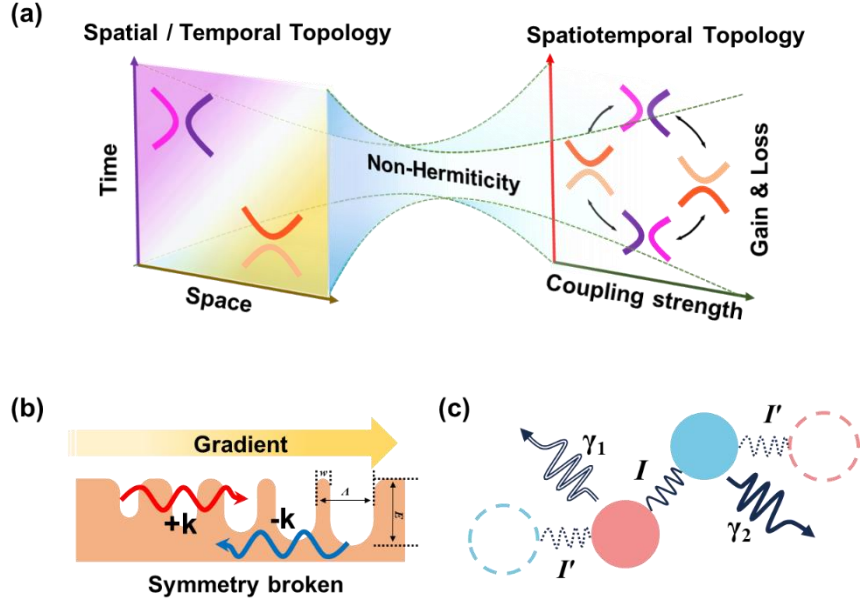

FIG. 1. Spatiotemporal topology in non-Hermitian system modulated by symmetry broken and unbroken. (**a**) Mapping spacetime through non-Hermiticity in Hamiltonian parameter space to achieve synthetic spacetime and spatiotemporal topology. (**b**) Waveguide-assisted grating structure enables tuning of loss and coupling via geometric parameters. (**c**) Schematics of SSH model incorporating loss.

Drawing on the fundamental symmetry between space and time, while energy-bands are typically functions of Bloch wavevector, the Hamiltonian for momentum-bands should be functions of frequency[14]. Using the transfer-matrix method (Supplemental Material S2), this Hamiltonian takes the form:

$$H_{\delta k}(\delta\omega) = \frac{-1}{1+\lambda^2}\begin{pmatrix} \delta\omega + i\frac{\eta}{2} + i\lambda m & -m - i\lambda\delta\omega - \lambda\frac{\eta}{2} \\ m + i\lambda\delta\omega - \lambda\frac{\eta}{2} & -\delta\omega + i\frac{\eta}{2} - i\lambda m \end{pmatrix}. \tag{3}$$

Conventional topological analysis methods like winding number in complex energy plane[31-33] or calculating Zak phase in generalized Brillouin zone[34-36] are no longer applicable for this Hamiltonian. This failure stems from two reasons. First, the eigenvalue spectrum resides in the complex-momentum plane rather than the complex-energy plane. Second, the absence of temporal periodicity in this static system prevents the frequency axis from being folded into a periodic Brillouin zone, so a closed integration contour cannot be defined. Therefore, we employ the biorthogonal parallel-transport method[23,44], which tracks the holonomy of Bloch eigenstates directly along an open path in the frequency parameter. We illustrate this approach by comparing two representative points in the parameter space, which are point 1 ($\lambda = 1$, $\eta = -0.5$) and point 2 ($\lambda = 1$, $\eta = 0.5$). For the left band ($\delta k < 0$) at these points, we follow the phase evolution of the normalized biorthogonal eigenstates as the frequency sweeps from the low-frequency limit ($\delta\omega \to -\infty$) to the high-frequency limits ($\delta\omega \to \infty$), during which the eigenvectors satisfy the biorthogonal parallel-transport condition (see details in Supplemental Material S3). While both points start with an initial phase of $\pi/2$ in low-frequency limit, their evolutionary paths diverge. As shown in Fig. 2b and Fig. S2, point 1 evolves toward a phase of 0, whereas point 2 approaches $\pi$ in the high-frequency limit. The resulting $\pi$ phase difference indicates that the two points belong to two distinct topological phases of the momentum-band[23].

Extending this analysis across the momentum-band region of the parameter space, we find that $\eta < 0$ and $\eta > 0$ correspond to two distinct topological phases labeled as phase I and phase II, respectively (Fig. 2a). The topological phase transition for momentum band happen to $\eta = 0$. For consistency, we apply the same parallel-transport gauge to the energy-band regions of III and IV to identify their topological phases. Along the line $\eta = 0$, the system reduces to a Hermitian limit with chiral symmetry, allowing the topological invariant to be evaluated using standard Hermitian methods (Supplemental Material S4). Corresponding phase boundary separates trivial and non-trivial phases whose transition remains at $\lambda = 1$ (Fig. 2c), consistent with the non-Hermitian phase diagram. This consistency allows the same topological labels to be extended across the full parameter space for energy-band regime. Consequently, the parameter space ($\lambda$, $\eta$) maps out a complete spatiotemporal topological phase diagram (Fig. 2a), capturing both energy-band and momentum-band regions, each with two distinct topological phases. The special point (1, 0) in this map serves as a spatiotemporal four-phase transition point where the boundaries of all four topological phases intersect. This diagram unifies previous separate phase-transition paths. The loss-induced momentum-band topology transition[23] can be marked by the green arrow. The energy-band transition in Ref. [17,43] governed by coupling strength can be labeled as the brown arrow path. The non-Hermitian SSH model in this work further supports a continuous spatiotemporal topological transition (e. g. paths ①-③ in Fig. 2a) across the phase diagram, bridging the two previously separate realms within a static parameter space.

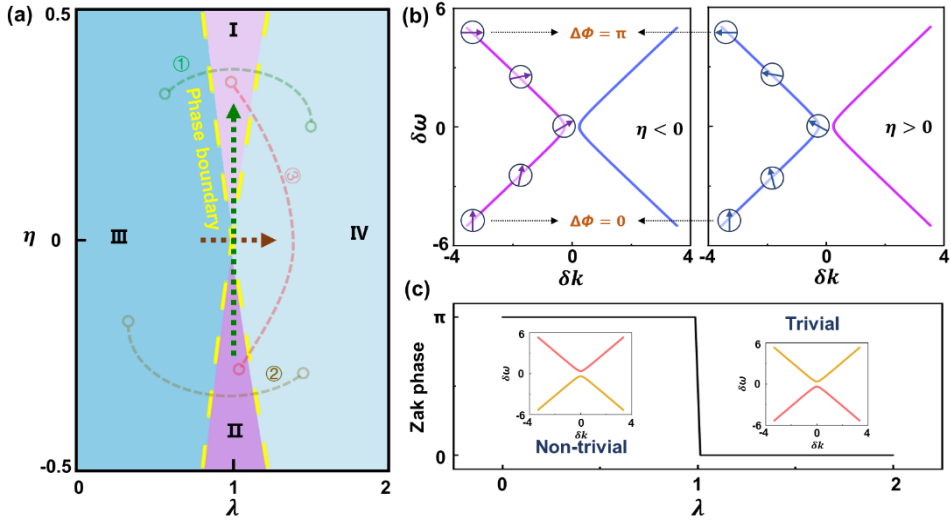

FIG. 2. The spatiotemporal topological transition in parameter space. (**a**) Phase diagram in the parameter space ($\lambda$, $\eta$). The regions labeled as phase I and II (purple) and phase III and IV (blue) correspond to energy- and momentum-band gaps, respectively. The phase boundary is indicated by the yellow dashed line. Other dashed lines indicate the phase trajectories. (**b**) Band structure for phases I and II. Arrows on the left bands show the phase in parallel transport. The final states accumulate a relative phase of π, signaling a topological transition. (**c**) Quantized Zak phase in the Hermitian system along the line of $\eta = 0$.

*Experimental observation of momentum gap* — We implement a one-dimensional grating–waveguide coupled structure (Fig. 1b) for the developed SSH model. Quasinormal-mode (QNM) method simulations is conducted in transverse electric (TE) polarization considering the dye absorption properties from dipole resonance. The results predict that a pair of EPs and a momentum-band gap should emerge under specific geometric parameters in which grating period $\Lambda$=395 nm, $E = 405$ nm and $D = 0.418$ (Fig. S5). To fabricate a structure that meets these conditions, we employed dual-beam interference lithography with a uniform exposure and a tailored development process (Fig. S3). The grating period $\Lambda$ was determined by the interference angle, while the DC and

ED were controlled by varying the exposure dose and leveraging the nonlinear response of the photoresist (see fabrication details and optical setup in Supplemental Material S5).

Figure 3a directly illustrates the experimental observation of the Bloch momentum-band gap through angle-resolved reflection spectroscopy in TE polarization, using a home-built high-resolution system (see Supplemental Material S7). The bandgap is centered at a wavelength of 603.6 nm, corresponding to $k = 0$. However, in transmission measurements, the bandgap is obscured by pronounced mode broadening due to non-Hermitian energy dissipation[17]. As shown in Fig. S7, while the reflection spectrum clearly reveals momentum-band gap splitting, corresponding in-situ transmission spectrum loses all band structure (Fig. S7b and S7d). To recover the band structure from the broadened transmission data, we developed a spectral-derivative method based on the strong wavelength sensitivity of the resonant absorption governed by the optical structure. Each eigenmode appears as a resonance peak in transmission, whose maximum satisfies $dT/d\lambda = 0$. Extracting these zero-derivative points effectively removes the influence of material absorption and mode broadening, yielding a reconstructed band structure with high spectral clarity as shown in Fig. 3b (details in Supplemental Material S8). The reconstructed experimental bands show excellent agreement with the theoretical structure derived from the non-Hermitian SSH model (white dashed line in Fig. 3b), confirming the momentum-band gap predicted by the Hamiltonian. The agreement between the reconstructed bands and the theoretical model identifies two EPs at ±0.202°, where the bands merge in their real part. Direct spectroscopic evidence for the EPs is provided by the reflection profiles at the marked incident angles of 0.202° and 0.542° (Fig. 3c). A single resonance peak at the EP splits into two distinct peaks as the angle increases, visually confirming the degeneracy at the EP and the subsequent band splitting.

Meanwhile, modulating the waveguide thickness in SPCs further introduce higher-order Bloch modes coupling both at and off the Γ point, establishing the conditions for multiple EP pairs. These resulting coupling profiles still remain captured by the established non-Hermitian SSH Hamiltonian with the coupling point extended to off-Γ point (Supplemental Material S2), validating the generality of our parameter space framework. The resulting simultaneous momentum-band gaps (Fig. 3d) provide direct experimental evidence for such multi-EP configurations, arising from coupled first- and second-order Bloch modes. Reflection profiles (Fig. S8) locate these EP pairs at distinct wavelengths of 613.3 nm, 623.5 nm, and 634.5 nm, corresponding to incidence angles of (-

0.33°, 0.33°), (-2.31°, 1.65°), (1.53°, 2.01°) and (-0.39°, 0.39°), respectively.

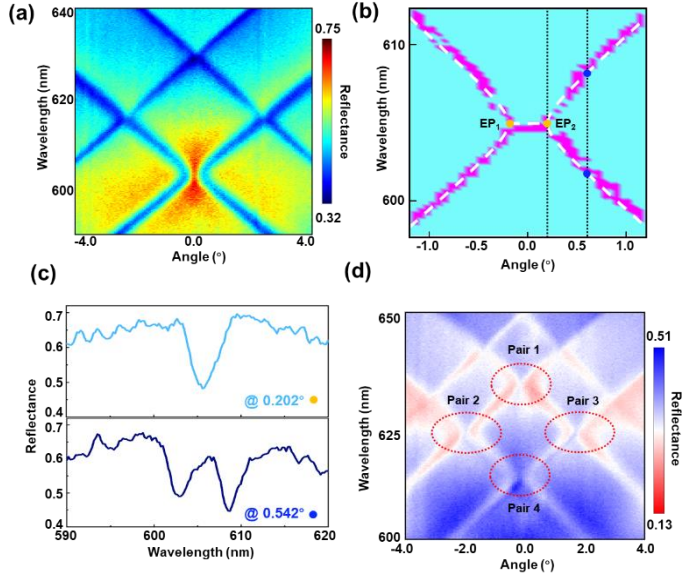

FIG. 3. Experimental demonstration of the Bloch momentum-band gap. (**a**) The experimental angle-resolved reflectance spectrum. (**b**) Reconstructed band dispersion relation derived from the angle-resolved transmittance spectra by spectral-derivative method, which agrees with the theoretical Hamiltonian marked by white dashed line. (**c**) Reflectance spectra at specific angles of EP and away from EP, as marked by the black dashed lines in **b**, proving the EP degeneracy and band splitting. (d) The emergence of multiple EP pairs.

*Spatiotemporal topological phase transition in gradient spatial PCs* — We further compute the band structure across a wide range of geometric parameters (ED and DC) using QNM simulations. The resulting phase diagram (Fig. 4a) plots the normalized band-gap widths. Crucially, fitting the simulated bands to the theoretical Hamiltonian shows that the energy band of structure with large DC is topologically non-trivial, whereas the band corresponding to small DC is trivial. The transition between these two distinct phases traverses a momentum-band gap and is accompanied by a band inversion of the symmetry-protected bound state in the continuum (SP-BIC). At the large DC, the upper band shows an odd-parity mode with high radiative quality factor ($Q_{\text{rad}}$), which is the characteristic of a SP-BIC[26,45,46], while the lower band is radiative with even parity (Fig. S9a-d). At the small DC, this ordering is inverted (Fig. S9e-h). This exchange of BIC between the two bands signals a genuine topological phase transition[17], in which the underlying topological order swaps between the upper and lower energy bands across the phase boundary.

To realize a continuous topological transition, we extend the design to gradient spatial PCs in which ED and DC vary continuously along the sample. This gradient serves two essential functions. First, it enables spatially resolved tuning of the relative coupling strength and relative loss, thereby projecting the abstract parameter space onto a real-space coordinate. Secondly, the spatial gradient renders the modulation experienced by waves propagating in opposite directions asymmetric in $k$-space. This wavevector-dependent asymmetry effectively breaks time-reversal symmetry along the propagation axis, akin to the non-reciprocal propagation in Faraday rotation. By introducing a synthetic arrow of time into the otherwise static system, this symmetry breaking becomes essential for topology dynamics on a spatial platform. Experimentally, the gradient was fabricated by introducing a Gaussian-shaped interference envelope into the dual-beam lithography process. This Gaussian envelope and nonlinear photoresist response produced a smooth gradient in DC and ED, with a smaller DC and deeper ED at the center, evolving into a larger DC and shallower ED toward the edges (details in Supplemental Material S11). Scanning electron microscopy (SEM) characterization confirms the realized gradient (Fig. S11). Corresponding $E$ decreases from 552.1 nm at the center to 337.5 nm at the edge, while $D$ concurrently increases from 0.327 to 0.490. The two marked points in Fig. 4a correspond to the geometric parameters at the center and edge of the fabricated sample. The white dashed line schematically represents the experimental spatial scanning trajectory, which switches between trivial and non-trivial energy-band phases while crossing momentum-band gap region.

We performed a continuous spatial scan across the sample and analyze the resulting angle-resolved spectra (Fig. S12a). The derivative of the normal-incidence reflectance $dR/d\lambda$ serves as a direct probe of the topological phase, evident in Fig. 4b. The sign of the derivative indicates whether the SP-BIC lies in the upper or lower band, while its magnitude drops to near zero in the momentum-gap regime (details in Supplemental Material S12). Applying this spectral-derivative analysis to the full scan yields a clear experimental phase map (Fig. 4c), which clearly separates the sample into three regions. The central region corresponds to a topologically trivial energy-band phase while the edge region indicates non-trivial energy-band phase. Between them is the momentum-band gap region. The spatial scan from the center (I) to the edge (III), crossing the intermediate momentum-gap region (II), directly maps onto path ① or ② in the theoretical phase diagram (Fig. 2a). Moreover, the two momentum band gap regions encountered along the scan lie at opposite gradient directions relative to time reversal symmetry breaking, leading to distinct topological characters. Therefore, another scan starting from one momentum gap (II), crossing the trivial energy-band region (I), and ending at the other momentum gap (IV) corresponds to path ③ in the phase diagram, effectively bridging two momentum-gap regions. Thus, by simply translating the probe across the static graded structure, we experimentally trace the key pathways predicted by the global spatiotemporal phase diagram. These results demonstrate that a single gradient PC platform can simultaneously host both energy-band and momentum-band gaps. Spatiotemporal topological switching is achieved in situ simply by translating the probe position across the sample (video, see Supplemental Material S13).

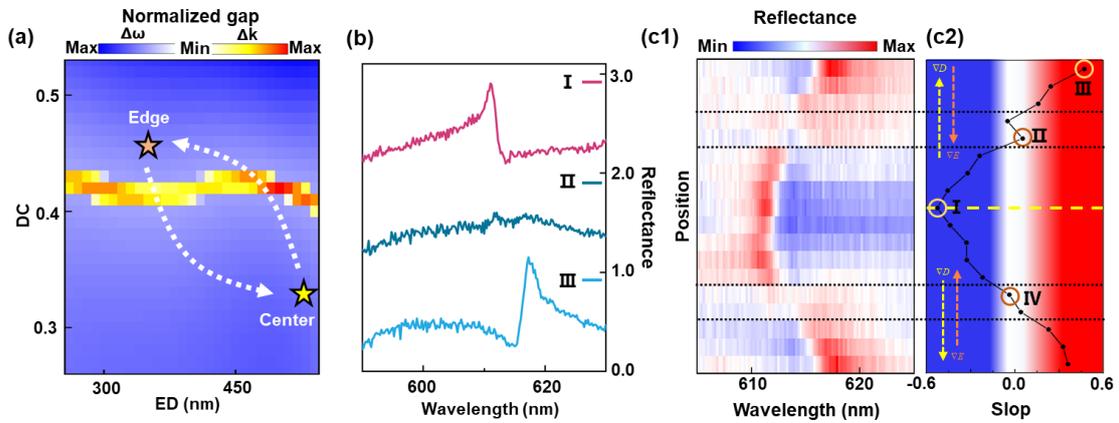

FIG. 4. Continuous spatiotemporal topological transition in fabricated gradient spatial PCs. (**a**) Simulated energy and momentum-band gap varies with DC and ED. White dashed arrows indicate the spatial scan path across the sample. (**b**) Normal-incidence reflectance spectra extracted at edge (I), medium (II), and center (III) of sample, corresponding to the trivial energy-band, the momentum-band, and non-trivial energy-band, respectively. (**c**) Position-dependent reflectance spectra at 0.0° from angle-resolved measurements (**c1**), and the corresponding derivative value curve obtained from the *dR/dλ* at each point (**c2**). The dashed arrows represent the gradient direction of the DC and ED, respectively. The structural parameters are mirror-symmetric about the yellow dashed line.

*Conclusion and discussions*— Based on a waveguide-assisted non-Hermitian SSH model, this work construct a global spatiotemporal topological phase map. This theoretical framework unifies energy-band and momentum-band topologies within a static parameter space, enabling continuous topological control across multiple phases. This advances beyond previous studies that treated spatial and temporal topologies separately, each limited to describing at most two phases. The developed topological characterization for non-chiral non-Hermitian systems is extendable to higher-dimensional and multiband settings. Experimentally, we projected this phase map in a spatially graded PC. Simply by translating the probe, we directly observe multiple EP pairs and a continuous spatiotemporal topological transition. The introduced spectral-derivative method resolves momentum gaps obscured by non-Hermitian broadening. These establish a parameter space mapping design principle offers a more accessible and cost-effective static platform for generating EPs and Bloch momentum-band gaps than conventional optical systems[16,47]. These advances open new avenues for topological photonics, robust signal control, and nonreciprocal meta-devices operating in static, loss-engineered environments.

*Acknowledgments*—This work was supported by the National Natural Science Foundation of China (Grant Nos. 92150109 and 61975018) and Beijing Key Laboratory of High-Entropy Energy materials and Devices, Beijing Institute of Nanoenergy and Nanosystems (No. GS2025ZD011).

The authors declare no conflicts of interest.

*Data availability*—The data that support the findings of this article are not publicly available. The data are available from the authors upon reasonable request.

[1] K. Vonklitzing, G. Dorda, and M. Pepper, NEW METHOD FOR HIGH-ACCURACY DETERMINATION OF THE FINE-STRUCTURE CONSTANT BASED ON QUANTIZED HALL RESISTANCE, Physical Review Letters **45**, 494 (1980).
[2] J. P. Eisenstein and H. L. Stormer, THE FRACTIONAL QUANTUM HALL-EFFECT, Science **248**, 1510 (1990).
[3] M. C. Rechtsman, J. M. Zeuner, Y. Plotnik, Y. Lumer, D. Podolsky, F. Dreisow, S. Nolte, M. Segev, and A. Szameit, Photonic Floquet topological insulators, Nature **496**, 196 (2013).
[4] Z. Wang, Y. D. Chong, J. D. Joannopoulos, and M. Soljacic, Observation of unidirectional backscattering-immune topological electromagnetic states, Nature **461**, 772 (2009).
[5] T. Ma and G. Shvets, All-Si valley-Hall photonic topological insulator, New J. Phys. **18**, 9, 025012 (2016).
[6] L. J. Maczewsky, J. M. Zeuner, S. Nolte, and A. Szameit, Observation of photonic anomalous Floquet topological insulators, Nature Communications **8**, 7, 13756 (2017).
[7] Y. Zeng, U. Chattopadhyay, B. Zhu, B. Qiang, J. Li, Y. Jin, L. Li, A. G. Davies, E. H. Linfield, B. Zhang, Y. Chong, and Q. J. Wang, Electrically pumped topological laser with valley edge modes, Nature **578**, 246 (2020).
[8] J. X. Chen, C. Y. Tang, A. Chen, J. H. Si, Z. Z. Yi, D. X. Xin, M. J. Wang, and W. H. Zheng, Self-

Organized Lasing of Delocalized State Enabled by Non-Hermitian Manipulation and Chiral Symmetry, Laser Photon. Rev., 9, e01772 (2025).

[9] X.-R. Mao, W.-J. Ji, S.-L. Wang, H.-Q. Liu, B. Wu, X.-J. Wang, L. Liu, L. Zhou, H. Ni, Z. Niu, and Z. Yuan, A single-photon source based on topological bulk cavity, Light: Science & Applications **14**, 295 (2025).

[10] E. Lustig, Y. Sharabi, and M. Segev, Topological aspects of photonic time crystals, Optica **5**, 1390 (2018).

[11] H. Moussa, G. Y. Xu, S. X. Yin, E. Galiffi, Y. Ra'di, and A. Alu, Observation of temporal reflection and broadband frequency translation at photonic time interfaces, Nat. Phys. **19**, 863 (2023).

[12] J. H. Dong, S. H. Zhang, H. He, H. A. Li, and J. J. Xu, Nonuniform Wave Momentum Band Gap in Biaxial Anisotropic Photonic Time Crystals, Physical Review Letters **134**, 7, 063801 (2025).

[13] E. Galiffi, R. Tirole, S. X. Yin, H. N. Li, S. Vezzoli, P. A. Huidobro, M. G. Silveirinha, R. Sapienza, A. Alù, and J. B. Pendry, Photonics of time-varying media, Adv. Photonics **4**, 32, 014002 (2022).

[14] Y. Sharabi, E. Lustig, A. Dikopoltsev, Y. Lumer, and M. Segev, Spatiotemporal photonic crystals, Optica **9**, 585 (2022).

[15] J. Xiong, X. D. Zhang, L. J. Duan, J. R. Wang, Y. Long, H. N. Hou, L. T. Yu, L. Y. Zou, and B. L. Zhang, Observation of wave amplification and temporal topological state in a non-synthetic photonic time crystal, Nature Communications **16**, 8, 11182 (2025).

[16] L. Wang, H. Liu, J. W. Liu, A. X. Liu, J. L. Huang, Q. N. Li, H. Dai, C. H. Zhang, J. B. Wu, K. B. Fan, H. B. Wang, B. B. Jin, J. Chen, and P. H. Wu, Photoswitchable exceptional points derived from bound states in the continuum, Light-Science & Applications **14**, 9, 377 (2025).

[17] W. Li, C. Luo, S. B. Tian, R. X. Zheng, G. Z. Geng, H. F. Yang, B. L. Liu, Q. H. Song, Y. Guo, and C. Z. Gu, Topological Band Engineering in q-BICs and EPs Derived from Visible Range Plasmons, Nano Lett. **25**, 6117 (2025).

[18] S. S. Tong, Q. C. Zhang, G. H. Li, K. Zhang, C. Xie, and C. Y. Qiu, Observation of momentum-band topology in PT-symmetric Floquet lattices, Nature Communications **16**, 8, 9975 (2025).

[19] C. T. Chan, Essay: Photonic Crystals as a Platform to Explore New Physics, Physical Review Letters **135**, 080001 (2025).

[20] Q. H. Song, M. Odeh, J. Zúñiga-Pérez, B. Kanté, and P. Genevet, Plasmonic topological metasurface by encircling an exceptional point, Science **373**, 1133 (2021).

[21] C. T. Chan, Essay: Photonic Crystals as a Platform to Explore New Physics, Physical Review Letters **135**, 9, 080001 (2025).

[22] A. Guo, G. J. Salamo, D. Duchesne, R. Morandotti, M. Volatier-Ravat, V. Aimez, G. A. Siviloglou, and D. N. Christodoulides, Observation of PT-Symmetry Breaking in Complex Optical Potentials, Physical Review Letters **103**, 4, 093902 (2009).

[23] M. W. Li, J. W. Liu, X. L. Wang, W. J. Chen, G. C. Ma, and J. W. Dong, Topological Temporal Boundary States in a Non-Hermitian Spatial Crystal, Physical Review Letters **135**, 8, 187101 (2025).

[24] S. S. Tong, Q. C. Zhang, L. J. Qi, G. H. Li, X. L. Feng, and C. Y. Qiu, Observation of Floquet-Bloch Braids in Non-Hermitian Spatiotemporal Lattices, Physical Review Letters **134**, 8, 126603 (2025).

[25] J. Feis, S. Weidemann, T. Sheppard, H. M. Price, and A. Szameit, Space-time-topological events in photonic quantum walks, Nature Photonics **19**, 518 (2025).

[26] H. Y. Yuan, J. Y. Liu, X. L. Wang, Z. M. Zeng, Q. W. Jia, J. W. Shi, D. H. Liu, and Z. N. Wang, Dynamically Switchable Polarization Lasing Between q-BIC and Bragg Resonance Modes, Laser Photon. Rev., 10 (2025).

[27] A. D. Li, H. Wei, M. Cotrufo, W. J. Chen, S. Mann, X. Ni, B. C. Xu, J. F. Chen, J. Wang, S. H. Fan, C. W. Qiu, A. Alú, and L. Chen, Exceptional points and non-Hermitian photonics at the nanoscale, Nature Nanotechnology **18**, 706 (2023).

[28] M. A. Miri and A. Alù, Exceptional points in optics and photonics, Science **363**, 42, eaar7709 (2019).

[29] S. K. Özdemir, S. Rotter, F. Nori, and L. Yang, Parity-time symmetry and exceptional points in photonics, Nature Materials **18**, 783 (2019).

[30] L. Feng, R. El-Ganainy, and L. Ge, Non-Hermitian photonics based on parity-time symmetry, Nature Photonics **11**, 752 (2017).

[31] K. Zhang, Z. S. Yang, and C. Fang, Correspondence between Winding Numbers and Skin Modes in Non-Hermitian Systems, Physical Review Letters **125**, 6, 126402 (2020).

[32] Y. K. Yang, H. Hu, L. L. Liu, Y. H. Yang, Y. X. Yu, Y. Long, X. Z. Zheng, Y. Luo, Z. Li, and F. J. Garcia-Vidal, Topologically Protected Edge States in Time Photonic Crystals with Chiral Symmetry, Acs Photonics **12**, 2389 (2025).

[33] C. Hou, L. F. Li, G. Z. Wu, Y. Ruan, S. H. Chen, and F. Baronio, Topological edge states in one-dimensional non-Hermitian Su-Schrieffer-Heeger systems of finite lattice size: Analytical solutions and exceptional points, Physical Review B **108**, 12, 085425 (2024).

[34] S. Y. Yao and Z. Wang, Edge States and Topological Invariants of Non-Hermitian Systems, Physical Review Letters **121**, 8, 086803 (2018).

[35] K. Yokomizo and S. Murakami, Non-Bloch Band Theory of Non-Hermitian Systems, Physical Review Letters **123**, 6, 066404 (2019).

[36] S. Tsubota, H. Yang, Y. Akagi, and H. Katsura, Symmetry-protected quantization of complex Berry phases in non-Hermitian many-body systems, Physical Review B **105**, 6, L201113 (2022).

[37] S. H. Chen, A. Basit, L. F. Li, C. Hou, Y. Ruan, Y. T. Wei, and Z. H. Ni, Non-Hermitian Topological Lattice Photonics: An Analytic Perspective, Adv. Photon. Res. **6**, 24 (2025).

[38] Z. Q. Jiao, S. Longhi, X. W. Wang, J. Gao, W. H. Zhou, Y. Wang, Y. X. Fu, L. Wang, R. J. Ren, L. F. Qiao, and X. M. Jin, Experimentally Detecting Quantized Zak Phases without Chiral Symmetry in Photonic Lattices, Physical Review Letters **127**, 6, 147401 (2021).

[39] Z. T. Wang, X. D. Wang, Z. C. Hu, D. Bongiovanni, D. Jukic, L. Q. Tang, D. H. Song, R. Morandotti, Z. G. Chen, and H. Buljan, Sub-symmetry-protected topological states, Nat. Phys. **19**, 992 (2023).

[40] S. Verma and M. J. Park, Non-Bloch band theory of subsymmetry-protected topological phases, Physical Review B **110**, 14, 035424 (2024).

[41] H. C. Po, H. Watanabe, and A. Vishwanath, Fragile Topology and Wannier Obstructions, Physical Review Letters **121**, 6, 126402 (2018).

[42] Y. A. Wang, H. X. Wang, L. Liang, W. W. Zhu, L. Z. Fan, Z. K. Lin, F. F. Li, X. Zhang, P. G. Luan, Y. Poo, J. H. Jiang, and G. Y. Guo, Hybrid topological photonic crystals, Nature Communications **14**, 9, 4457 (2023).

[43] Z. T. Wang, D. Bongiovanni, X. D. Wang, Z. C. Hu, D. Jukic, D. H. Song, J. J. Xu, R. Morandotti, Z. G. Chen, and H. Buljan, Hidden multi-topological phases mediated by constrained inter-cell coupling, Elight **6**, 13, 2 (2026).

[44] R. Resta, Manifestations of Berry's phase in molecules and condensed matter, J. Phys.-Condes. Matter **12**, R107 (2000).

[45] C. W. Hsu, B. Zhen, J. Lee, S. L. Chua, S. G. Johnson, J. D. Joannopoulos, and M. Soljacic, Observation of trapped light within the radiation continuum, Nature **499**, 188 (2013).

[46] F. Cao, M. Zhou, C.-W. Cheng, H. Li, Q. Jia, A. Jiang, B. Lyu, D. Liu, D. Han, S. Gwo, and J. Shi,

Interaction of plasmonic bound states in the continuum, Photonics Research **11**, 724 (2023).

[47] B. Zhen, C. W. Hsu, Y. Igarashi, L. Lu, I. Kaminer, A. Pick, S. L. Chua, J. D. Joannopoulos, and M. Soljacic, Spawning rings of exceptional points out of Dirac cones, Nature **525**, 354 (2015).